\documentclass[aps,prl,a4paper,preprintnumbers,tightenlines,superscriptaddress,twocolumn,floatfix,showpacs,final]{revtex4-1}
\usepackage{amsmath}
\usepackage{amssymb}
\usepackage{float}
\usepackage[dvips]{graphics}
\usepackage[pdftex]{graphicx}

\newcommand{\be}{\begin{eqnarray}}
\newcommand{\ee}{\end{eqnarray}}

\newcommand{\tr}{{\rm tr}\,}

\begin{document}

\title{A New Solution of the Cosmological Constant Problems}
\author{John D. Barrow}
\author{Douglas J. Shaw}
\affiliation{DAMTP, Centre for Mathematical Sciences, Cambridge CB3 0WA,
United Kingdom}
\date{\today}

\begin{abstract}
We extend the usual gravitational action principle by promoting the bare
cosmological constant (CC) from a parameter to a field which can take many
possible values. Variation leads to a new integral constraint equation which
determines the classical value of the effective CC that dominates the wave
function of the universe. In a realistic cosmological model, the expected
value of the effective CC, is calculated from measurable quantities to be $%
O(t_{\mathrm{U}}^{-2})$,  as observed, where $t_{\mathrm{U}}$ is the present
age of the universe in Planck units,. Any application of our model produces
a falsifiable prediction for $\Lambda $ in terms of other measurable
quantities. This leads to a specific falsifiable prediction for the observed
spatial curvature parameter of $\Omega _{k0}=-0.0055$. Our testable proposal
requires no fine tunings or extra dark-energy fields but does suggest a new view of time and cosmological evolution.
\end{abstract}

\maketitle

The cosmological constant (CC) was introduced by Einstein in 1917 to ensure
that general relativity (GR) admitted a static cosmological solution.
Introducing a CC, $\lambda $, required the addition of a term $-\lambda
g_{\mu \nu }$ to the original field equations: 
\begin{equation*}
G^{\mu \nu }=\kappa \left\langle T^{\mu \nu }\right\rangle \rightarrow
G^{\mu \nu }=\kappa \left\langle T^{\mu \nu }\right\rangle -\lambda g^{\mu
\nu },
\end{equation*}%
where $G^{\mu \nu }=R^{\mu \nu }-Rg^{\mu \nu }/2$, $R_{\mu \nu }$ is the
Ricci curvature of $g_{\mu \nu }$, and $\left\langle T^{\mu \nu
}\right\rangle $ is the expected energy-momentum tensor of matter; $\kappa
=8\pi G$, $c=\hbar =1$.  The discovery that the universe was expanding
removed Einstein's original motivation for $\lambda $ but it was later
appreciated that there were other, more fundamental, reasons for its
presence. Quantum fluctuations result in a vacuum energy, $\rho _{\mathrm{vac%
}},$ that contributes to the $\left\langle T^{\mu \nu }\right\rangle $ 
\begin{equation*}
\left\langle T^{\mu \nu }\right\rangle =T_{\mathrm{m}}^{\mu \nu }-\rho _{%
\mathrm{vac}}g^{\mu \nu },
\end{equation*}%
where $T_{\mathrm{m}}^{\mu \nu }$ vanishes \emph{in vacuo} and hence 
\begin{equation*}
G^{\mu \nu }=\kappa T_{\mathrm{m}}^{\mu \nu }-\Lambda g^{\mu \nu },\qquad
\Lambda =\lambda +\kappa \rho _{\mathrm{vac}}.
\end{equation*}%
The vacuum energy contributes $\kappa \rho _{\mathrm{vac}}$ to the effective
CC, $\Lambda $. Even if the `bare' CC, $\lambda $, is assumed to vanish, the
effective CC will generally be non-zero. For $\Lambda =0$, the $\lambda $,
and $\kappa \rho _{\mathrm{vac}}$ terms must exactly cancel. Since there is
no a priori link between the values of $\lambda $ and $\kappa \rho _{\mathrm{%
vac}}$ this seems improbable. Without such a cancellation, it is natural
that $|\Lambda |\gtrsim O(\kappa \rho _{\mathrm{vac}})$. 
% Measurements of the Casimir force show that $\rho_{\rm vac}$, or at least changes in it, is physically real.

At late cosmic times $\rho _{\mathrm{vac}}$ does not evolve. Given the
standard model of particle physics, and reasonable (e.g. supersymmetric)
extensions of it, a late-time $\rho _{\mathrm{vac}}$ of at least $M_{\mathrm{%
EW}}^{4}$ $\sim (246\,\mathrm{GeV)}^{4}$ appears to be unavoidable. Hence,
it seems natural that $\rho _{\mathrm{vac}}^{\mathrm{eff}}=\kappa
^{-1}\Lambda \gtrsim M_{\mathrm{EW}}^{4}$. This \emph{cannot} be the case
because measurements of the expansion rate give $\rho _{\mathrm{vac}}^{%
\mathrm{eff}}\approx (2.4\times 10^{-12}\,\mathrm{GeV)}^{4}$ \cite%
{Komatsu:2010fb}, at least $10^{56}$ times smaller than the expected quantum
contribution. This is the \emph{cosmological constant problem}.\
Equivalently, assuming the estimate of $\rho _{\mathrm{vac}}$ from quantum
fluctuations is accurate we ask why $\lambda \approx -\kappa \rho _{\mathrm{%
vac}}$ to at least $56$ decimal places? Furthermore, the time $t_{\Lambda
}=\Lambda ^{-1/2}\approx 9.7\,\mathrm{Gyrs}$ is curiously close to the
present age of the universe, $t_{U}\approx 13.7\,\mathrm{Gyrs}$. First
Barrow and Tipler \cite{btip}, and then Efstathiou \cite{ef} and Weinberg 
\cite{sw}, derived anthropic upper limits on $|\Lambda |$ by requiring that
inhomogeneities grow by gravitational instability long enough for galaxies
to form. For $\Lambda >0$ this requires $t_{\Lambda }\gtrsim 0.7\,\mathrm{%
Gyrs}$. However, there is still no reason why the apparently fixed time, $%
t_{\Lambda },$ should correlate with an observer-dependent time-scale such
as $t_{\mathrm{U}}$. This is the \emph{coincidence problem}.

We propose a simple extension of the usual action principle in which the
bare CC, $\lambda $, will be promoted from a parameter to a `field'. The
variation leads to a new field equation which determines the value of $%
\lambda ,$ and hence the effective CC, in terms of other properties of the
observed universe. Crucially, one finds that the observed classical history
naturally has $t_{\Lambda }\sim t_{\mathrm{U}}$. Fuller details are
presented elsewhere \cite{longpaper}. When it is applied to GR, $\lambda $
(and hence $\Lambda $ except when $\rho _{\mathrm{vac}}$ evolves due to,
say, a phase transition) is a true constant and is \emph{not} seen to
evolve. Hence, the resulting history is indistinguishable from GR with the
value of $\Lambda $ put in by hand. Nonetheless, for given theory of gravity
such as GR, our model produces a firm prediction for $\Lambda $ in terms of
other measurable quantities and is testable by future observations. It
should be stressed that our proposal is equally applicable to theories of
gravity other than GR and to theories with more than 4 spacetime dimensions.
As in 4-d GR, $t_{\Lambda }$ is still expected to be $O(t_{\mathrm{U}})$.

If our model is correct, assuming an (approximately) homogeneous and
isotropic GR cosmology, the measured value of $\Lambda$ requires a specific
value for the dimensionless spatial curvature, $\Omega _{k0}$, of the
observable universe. The predicted $\Omega _{k0}$ is consistent with current
observational limits and large enough to be detected in the near future. Our
model also specifies the probability, $f(\Lambda )\,\mathrm{d}\Lambda $
observing a CC in the range $[\Lambda ,\Lambda +\,\mathrm{d}\Lambda ]$.
Crucially, $f(\Lambda )$ is independent of the prior weighting given to
different values of $\Lambda $ in the wave function of the universe. We find
that the observed value of $\Lambda $ is indeed typical, as is a coincidence
between $t_{\Lambda }$ and $t_{\mathrm{U}}$. Our proposal provides a
realistic and falsifiable model of the universe that avoids the CC and
coincidence problems.

Define the total action of the universe on a manifold $\mathcal{M}$ with
boundary $\partial \mathcal{M}$ and effective CC $\Lambda ,$ matter fields $%
\Psi ^{a},$ and metric $g_{\mu \nu }$, to be $I_{\mathrm{tot}}[g_{\mu \nu
},\Psi ^{a},\Lambda ;\mathcal{M}]$. Usually, $\lambda $ is a fixed parameter
and the wave (partition) function of the universe, $Z[\lambda ;\mathcal{M}%
]\equiv Z_{\Lambda }[\mathcal{M}]$, is given by: 
\begin{equation*}
Z_{\Lambda }[\mathcal{M}]=\sum_{{}}e^{iI_{\mathrm{tot}}}\left[ \times \,\,%
\mathrm{gauge\,\,fixing\,\,terms}\right] ,
\end{equation*}%
where $\left\{ Q^{A}\right\} $ are some fixed boundary quantities
(generalized `charges') on $\partial \mathcal{M}$, and the sum is over all
histories (i.e. configurations of the metric and matter, $g_{\mu \nu },\Psi
^{a}$) consistent with these fixed charges. The dominant contribution to $%
Z_{\Lambda }[\mathcal{M}]$ is from the histories for which $I_{\mathrm{tot}}$
is stationary for $g_{\mu \nu }$ and $\Psi ^{a}$ variations that preserve
the $\left\{ Q^{A}\right\} $. In these dominant histories, the matter and
metric fields obey their classical field equations.

When the surface terms in the gravitational action are chosen to make $I_{%
\mathrm{tot}}$ first order in derivatives of the metric, for a non-null $%
\partial \mathcal{M}$ with induced 3-metric $\gamma _{\mu \nu }$, a small
general metric variation gives 
\begin{equation*}
2\kappa \delta I_{\mathrm{tot}}=\int_{\partial \mathcal{M}}|\gamma |^{\frac{1%
}{2}}\,\mathrm{d}^{3}x\,N^{\mu \nu }\delta \gamma _{\mu \nu }+\int_{\mathcal{%
M}}|g|^{\frac{1}{2}}\,\mathrm{d}^{4}x\,E^{\mu \nu }\delta g_{\mu \nu }.
\end{equation*}%
Put $g_{\mu \nu }=\bar{g}_{\mu \nu }+\delta g_{\mu \nu }^{(\mathcal{M})}$, $%
\bar{g}_{\mu \nu }=g_{\mu \nu }^{(0)}+\delta g_{\mu \nu }^{(\partial 
\mathcal{M})},$ where the $\delta g_{\mu \nu }^{(\mathcal{M})}$ vanish on $%
\partial \mathcal{M}$ but $\delta g_{\mu \nu }^{(\partial \mathcal{M})}$ do
not. The vanishing of $\delta I_{\mathrm{tot}}$ in $\mathcal{M}$ implies
that $E^{\mu \nu }[g_{\mu \nu }^{(0)}]=E^{\mu \nu }[\bar{g}_{\mu \nu }]=0$.
The classical field equations for the metric are $E^{\mu \nu }=0$ . The
variation $\delta I_{\mathrm{tot}}=0$ then requires that $\gamma _{\mu \nu }$
be fixed on $\partial \mathcal{M}$. However, if some part, $\partial 
\mathcal{M}_{u}$, of $\partial \mathcal{M}$ lies in the causal future of
another part, $\partial \mathcal{M}_{I}$, the choice of fixed $\gamma _{\mu
\nu }$ is constrained by $E^{\mu \nu }=0$. In this example, we define $%
\left\{ Q^{A}\right\} $ to be the smallest data set on $\partial \mathcal{M}$
that can be freely specified which, when combined with $E^{\mu \nu }=0$,
fixes $\gamma _{\mu \nu }$ up to a gauge choice on $\partial \mathcal{M}$.
This definition is then extended to the matter sector (for which the
classical field equations are $\Phi _{a}=0$). This is just a restatement of
the usual variational principle allowing for a causally interconnected $%
\partial \mathcal{M}$. Since $E^{\mu \nu }=0$ depends on $\Lambda $, fixed $%
\left\{ Q^{A}\right\} $ and $E^{\mu \nu }=\Phi _{a}=0$ only fixes $\gamma
_{\mu \nu }$ and boundary matter fields for given $\Lambda $, and we have $%
\left. \delta \gamma _{\mu \nu }\right\vert _{\partial \mathcal{M}}=\mathcal{%
H}_{\mu \nu }\delta \Lambda $ and $\left. \delta \Psi ^{a}\right\vert
_{\partial \mathcal{M}}=\mathcal{P}^{a}\delta \Lambda ,$which define $%
\mathcal{H}_{\mu \nu }$ and $\mathcal{P}^{a}$.

Our proposal for solving the CC problems is simply to promote the bare
cosmological constant, $\lambda $, from a fixed parameter to a field (albeit
one that is constant in space and time). A similar promotion of $\lambda $
occurs in studies of unimodular gravity. Equally, this promotion can arise
in a fundamental theory, e.g. string theory, where there are many distinct
vacua each with different minima of the vacuum energy.

The wave function of the universe, $Z[\mathcal{M}],$ now includes a sum over
all possible values of $\lambda $ in addition to the usual sum over
configurations of $g_{\mu \nu }$ and $\Psi ^{a}$ \cite{footnote1}. The
effective CC, $\Lambda $, is equal to $\lambda +\mathrm{const}$ and so a sum
over all possible values of $\lambda $ is equivalent to a sum over all $%
\Lambda $ and so 
\begin{equation*}
Z[\mathcal{M}]=\sum_{\lambda }\mu \lbrack \lambda ]Z[\lambda ;\mathcal{M}%
]=\sum_{\Lambda }\mu \lbrack \lambda ]Z_{\Lambda }[\mathcal{M}].
\end{equation*}%
where $\mu \lbrack \lambda ]$ is some unknown prior weighting on the
different values of $\lambda $. Provided $\mu \lbrack \lambda ]$ is not
strongly peaked at a particular $\lambda $-value, we find that (at least
classically) our model is independent of the choice of $\mu $. The classical
histories that dominate the wave function are those for which, with fixed $%
\left\{ Q^{A}\right\} $, $\delta I_{\mathrm{tot}}=0$ for variations in the
summed-over fields. For variations of $g_{\mu \nu }$ and $\Psi ^{a},$ this
gives $E^{\mu \nu }=\Phi _{a}=0$ as before. Since $\lambda $ is summed over,
a stationary $I_{\mathrm{tot}}$ also now requires $\delta I_{\mathrm{tot}%
}/\delta \lambda =\delta I_{\mathrm{tot}}/\delta \Lambda =0$.

We define $I_{\mathrm{class}}(\Lambda ;\mathcal{M})$ to be $I_{\mathrm{tot}}$
evaluated at the classical solution for $g_{\mu \nu }$ and $\Psi ^{a}$ and
fixed $\left\{ Q^{A}\right\} $; $\delta I_{\mathrm{tot}}/\delta \Lambda =0$
is then equivalent to 
\begin{equation}
\frac{\,\mathrm{d}I_{\mathrm{class}}(\Lambda ;\mathcal{M})}{\,\mathrm{d}%
\Lambda }=0,  \label{eq:phi:simp}
\end{equation}%
Eq.(\ref{eq:phi:simp}) yields a field equation for determining the classical
value of the effective CC. An observer sees a classical history with
effective CC, $\Lambda $, which satisfies Eq.(\ref{eq:phi:simp}). Since $%
\lambda $ is a true space-time constant, the effective CC will not be seen
to evolve in this classical history.

The solutions of Eq.(\ref{eq:phi:simp}) depend on the definition of $\mathcal{M}$%
, fixed $\left\{ Q^{A}\right\} $ and surface terms in $I_{\mathrm{tot}}$;
these choices should be well-motivated and consistent with the symmetries of
nature. We demand that all observables including $\Lambda $ should  be
influenced only by parts of the universe causally connected to the observer.
As Eq.(\ref{eq:phi:simp}) involves integrals over $\mathcal{M}$ and $\partial 
\mathcal{M}$, the only coordinate independent choice consistent with this
demand is that $\mathcal{M}$ is the observer's causal past. If our model's predictions are accurate, this requirement could indicate that a notion of causal order is a fundamental rather than emergent property of quantum space-time. The wave
function, $Z[\mathcal{M}]$, is then a sum over all possible configurations
in the causal past, and $\partial \mathcal{M}$ is composed of the observer's
past-light cone, $\partial \mathcal{M}_{u}$, and initial spacelike
singularity $\partial \mathcal{M}_{I}$, (where say $\tau =0$) \cite%
{blackholenote}. As we move towards $\partial \mathcal{M}_{I}$, the CC has
less and less influence on the evolution of the universe. This motivates
specifying the $\left\{ Q^{A}\right\} $ so that the initial state on $%
\partial \mathcal{M}_{I}$ is fixed independently of $\lambda $. On $\partial 
\mathcal{M}_{u}$, the fields then depend on $\Lambda $ through the classical
field equations in a calculable fashion. The canonical surface term choice
is the minimal term that renders the total action first order in metric and
matter derivatives \cite{footnoteINITIAL}. These choices are well-motivated
and natural; indeed there were no obvious and well-motivated alternatives.

There is now a simple argument for why $t_{\Lambda }\sim t_{\mathrm{U}}$ is
natural in our model. Schematically, with $I_{\mathrm{tot}}$ at most first
order in metric and matter derivatives, Eq.(\ref{eq:phi:simp}), is
equivalent to 
\begin{equation}
\int_{\mathcal{M}}|g|^{\frac{1}{2}}\,\mathrm{d}^{4}x=\frac{1}{2}%
\int_{\partial \mathcal{M}}|\gamma |^{\frac{1}{2}}\left[ N^{\mu \nu }%
\mathcal{H}_{\mu \nu }+\Sigma _{a}\mathcal{P}^{a}\right] \,\mathrm{d}^{3}x.
\label{eq:phi}
\end{equation}%
The left-hand side is just the 4-volume, $V_{\mathcal{M}}$, of $\mathcal{M}$%
. The right-hand side is a `holographic' term defined on the boundary (of
area $A_{\partial \mathcal{M}},$ say). Cosmologically $N^{\mu \nu }\mathcal{H}_{\mu \nu
} + \Sigma _{a}\mathcal{P}^{a} \sim O(\tr N/\Lambda) \sim O(H/\Lambda)$ where $H$ is
the Hubble constant (with $H(t_{U})\equiv $ $H_{0}$ today). Hence the
right-hand side of Eq.(\ref{eq:phi}) is $O(\Lambda ^{-1}H_{0}A_{\partial 
\mathcal{M}})$. So, we expect solutions of Eq.(\ref{eq:phi}) to have $%
\Lambda \sim O(H_{0})A_{\partial \mathcal{M}}/V_{\mathcal{M}}$. Typically, $%
H_{0}\sim A_{\partial \mathcal{M}}/V_{\mathcal{M}}$ and $H_{0}^{-1}$ is
determined by $t_{\Lambda }=\Lambda ^{-1/2}$ and the age of the universe $t_{%
\mathrm{U}}$. Eq.(\ref{eq:phi}) links the values of $t_{\Lambda }$ and $t_{%
\mathrm{U}}$ and, in the absence of fine-tunings, we naturally expect $%
t_{\Lambda }\sim O(t_{\mathrm{U}})$ and hence $\Lambda \sim O(1)t_{U}^{-2}$ (%
$\sim 10^{-122}$ in units where $G\equiv 1$). If there are extra dimensions
with volume $V_{D}$, then $A_{\partial \mathcal{M}}$ and $V_{\mathcal{M}}$
would both be multiplied by $V_{D}$ leaving $A_{\partial \mathcal{M}}/V_{%
\mathcal{M}}$ and the expectation $\Lambda \sim t_{U}^{-2}$ is unaltered 
\cite{sorkin}. If Eq.(\ref{eq:phi}) admits a classical solution, then the
classical value of the effective CC will have the observed magnitude, $%
O(t_{U}^{-2})\sim 10^{-122},$ without fine-tuning.

We now apply our model to our universe where gravity is described by GR to a
good approximation. The observed CC is given by the requirement that the
total action $I_{\mathrm{cl}}$ be stationary with respect to small changes
in $\lambda $, i.e. Eq.(\ref{eq:phi:simp}. We expand this equation by first
evaluating $I_{\mathrm{cl}}$ as a implicit function of $\lambda $. $I_{%
\mathrm{cl}}$ is the total action $I_{\mathrm{tot}}$ modulo the matter and
metric field equations, with 
\begin{equation*}
I_{\mathrm{tot}}=I_{\mathrm{EH}}+I_{\mathrm{CC}}+I_{\mathrm{GHY}}^{(u)}+I_{%
\mathrm{m}}+\dots ,
\end{equation*}%
where the $\dots $ represent the $\lambda $-independent surface terms on $%
\partial M_{\mathrm{I}}$. $I_{\mathrm{EH}}$ is the usual Einstein-Hilbert
action i.e. the integral of $(2\kappa )^{-1}\sqrt{-g}R$ over $\mathcal{M}$; $%
I_{\mathrm{CC}}$ and $I_{\mathrm{m}}$ are the cosmological constant and
matter actions respectively and $I_{\mathrm{GHY}}^{(u)}$ is the standard
Gibbons-Hawking-York surface term on $\partial \mathcal{M}_{u}$. We remove
the quantum vacuum energy from $I_{\mathrm{m}}$ and absorb it into the
effective CC, $\Lambda =\lambda +\kappa \rho _{\mathrm{vac}}$. $I_{\mathrm{CC%
}}$ and $I_{\mathrm{m}}$ are then the  the integrals of $-\kappa ^{-1}\sqrt{%
-g}\Lambda $ and $\sqrt{-g}\mathcal{L}_{\mathrm{m}}$ over $\mathcal{M}$
respectively. $\mathcal{L}_{\mathrm{m}}$ is the effective matter Lagrangian
density defined to vanish in vacuo; $T_{\mathrm{m}}^{\mu \nu }$ is the
associated energy-momentum tensor. Einstein's equations give $(2\kappa
)^{-1}R=2\kappa ^{-1}\Lambda -T_{\mathrm{m}}/2$ which we substitute into $I_{%
\mathrm{EH}}$. $I_{\mathrm{GHY}}^{(u)}$ can be transformed so that $I_{%
\mathrm{tot}}$ and $I_{\mathrm{cl}}$ can be written as a volume integral on $%
\mathcal{M}$ (see Ref. \cite{longpaper} for details).

For simplicity we focus on a homogeneous and isotropic cosmology with
metric: 
\begin{equation*}
\,\mathrm{d}s^{2}=a^{2}(\tau )\left[ -\,\mathrm{d}\tau
^{2}+(1+kx^{2}/4)^{-2}\,\mathrm{d}x^{i}\,\mathrm{d}x^{i}\right] ,
\end{equation*}%
where $k$ determines the spatial curvature. The observer is at $(\tau
,x)=(\tau _{0},0)$ and $\partial \mathcal{M}_{I}$ is the surface $\tau =0$
where $a=0$. We take $T_{\mathrm{m}}^{\mu \nu }=(\rho _{\mathrm{m}}+P_{%
\mathrm{m}})U^{\mu }U^{\nu }+P_{\mathrm{m}}g^{\mu \nu }$; $U^{\mu
}=-a^{-1}\nabla ^{\mu }\tau $. With $H=a_{,\tau }/a^{2}$, Einstein's
equations give $H^{2}=\kappa \rho _{\mathrm{m}}/3+\Lambda /3-k/a^{2}$ and $%
a^{-1}\rho _{\mathrm{m},\tau }=-3H(\rho _{\mathrm{m}}+P_{\mathrm{m}})$. We
find that to linear order in $O(kx^{2})$, $I_{\mathrm{cl}}$ is \cite%
{longpaper}: 
\begin{equation*}
I_{\mathrm{cl}}=\frac{4\pi }{3}\int_{0}^{\tau _{0}}a^{4}(\tau )(\tau
_{0}-\tau )^{3}\left[ \kappa ^{-1}\Gamma -P_{\mathrm{eff}}(a)\right] \,%
\mathrm{d}\tau .
\end{equation*}%
where $P_{\mathrm{eff}}=P_{\mathrm{m}}-\mathcal{L}_{\mathrm{m}}$ and $\Gamma
=(k/a^{2})[2/3+\tau /(\tau _{0}-\tau )]$. Contributions to $P_{\mathrm{eff}}$
can come from radiation, dark matter and baryonic matter (labelled `rad',
`dm' and `b' respectively). For radiation and dark matter, $P_{\mathrm{rad}%
}=\rho _{\mathrm{rad}}/3$, $\mathcal{L}_{\mathrm{rad}}/\rho _{\mathrm{rad}%
}\approx 0$ and $P_{\mathrm{dm}}/\rho _{\mathrm{dm}},\,\mathcal{L}_{\mathrm{%
dm}}/\rho _{\mathrm{dm}}\approx 0$. For baryonic matter, $P_{\mathrm{b}%
}/\rho _{\mathrm{b}}\approx 0$, $\mathcal{L}_{\mathrm{b}}=-\zeta _{\mathrm{b}%
}\rho _{\mathrm{b}},$ where for some $\zeta _{\mathrm{b}}\sim O(1)$ is
calculable in principle from QCD. The chiral bag model for baryon structure
gives the estimate $\zeta _{\mathrm{b}}\approx 1/2$ \cite{longpaper}. Since $%
\rho _{\mathrm{b}}\gg \rho _{\mathrm{rad}}$, the dominant contribution to $%
P_{\mathrm{eff}}$ comes from baryonic matter and $P_{\mathrm{eff}}\approx
\zeta _{\mathrm{b}}\rho _{\mathrm{b}}$. The terms in $I_{\mathrm{cl}}$ only
depend on $\lambda $ through the scale factor $a(\tau )$. We define $\delta
\ln a/\delta \lambda =\mathcal{A}(\tau )$. $\Gamma \propto a^{-2}$ and $P_{%
\mathrm{eff}}\approx \zeta _{\mathrm{b}}\rho _{\mathrm{b}}\propto 1/a^{3}$,
so $\delta (a^{4}\Gamma )/\delta \lambda =2\Gamma \mathcal{A}(\tau )$ and $%
\delta (a^{4}P_{\mathrm{eff}})/\delta \lambda \approx \zeta _{\mathrm{b}%
}\rho _{\mathrm{b}}\mathcal{A}(\tau )$; $\mathcal{A}(\tau )$ follows from
perturbing Einstein's equations with respect to $\Lambda $ and using $\delta
\ln a/\delta \Lambda =0$ initially. We\ find \cite{longpaper}: 
\begin{equation*}
\mathcal{A}(\tau )=\frac{a(\tau )H(\tau )}{6}\int_{0}^{\tau }\frac{\,\mathrm{%
d}\tau ^{\ast }}{H^{2}(\tau ^{\ast })}.
\end{equation*}%
Varying $I_{\mathrm{cl}}$ with respect to $\lambda ,$ we find that Eq.(\ref%
{eq:phi}) for the CC is equivalent to: 
\begin{equation}
k=\frac{\kappa \int_{0}^{\tau _{0}}(\tau _{0}-\tau )^{3}a^{4}\zeta _{\mathrm{%
b}}\rho _{\mathrm{b}}\mathcal{A}(\tau )\,\mathrm{d}\tau }{\int_{0}^{\tau
_{0}}a^{2}(\tau )(\tau _{0}-\tau )^{2}(4(\tau _{0}-\tau )+6\tau )\mathcal{A}%
(\tau )\,\mathrm{d}\tau }.  \label{eq:const}
\end{equation}
Note that this $k$ is the average spatial curvature in the causal past rather than necessarily the average spatial curvature of the whole space-time; hence $k > 0$ does \emph{not} require the universe to have a closed topology. 

Eq.(\ref{eq:const}) is a consistency condition that relates the value of $k$
to $\Omega _{\mathrm{b0}}=\kappa \rho _{\mathrm{baryon}}(\tau
_{0})/3H_{0}^{2}$, the observation time $\tau _{0}$ and, through $a(\tau )$
and $\mathcal{A}(\tau ),$ to $\Lambda $. So it gives $k=k_{0}(\Lambda ;\tau
_{0})$ and hence $\Lambda =\Lambda _{0}(k;\tau _{0})$. If our model is
valid, a measurement of $\Lambda $ at a given time predicts a specific value
of $k$ and hence $\Omega _{\mathrm{k0}}=-k/a_{0}^{2}H_{0}^{2}$. There are no
free parameters in this prediction. Eq.(\ref{eq:const}) requires $k>0$ i.e.
the observable universe has a positive spatial curvature. For our universe,
taking $\Omega _{\Lambda 0}\approx 0.73$, $\Omega _{\mathrm{b0}}\approx
0.0423$ and $T_{\mathrm{CMB}}=2.725\,\mathrm{K}$ we predict: 
\begin{equation*}
\Omega _{\mathrm{k0}}=-0.0055(2\zeta _{\mathrm{b}}).
\end{equation*}%
This is consistent (for all $\zeta _{\mathrm{b}}\in (0,1]$) with the current
95\% CI of $\Omega _{\mathrm{k0}}\in (-0.0133,0.0084)$ \cite{Komatsu:2010fb}%
. A combination of data from the current Planck satellite CMB survey with
measurements of baryon acoustic oscillations (BAO) will be able to test this
prediction of $\Omega _{\mathrm{k0}}$. 

Inflation in the early universe is usually invoked to explain why$\ $the
curvature term is so small today. The duration of inflation, given by the
number of e-folds $N$, depends on initial conditions since different
inflating regions in the same universe will have different $N$. Hence, $%
\Omega _{k}$ is an environmental parameter which is stochastically different
in each inflating region. In our model the extent to which the observed
value, $\Lambda _{\mathrm{obs}}$, of the CC is natural is determined by the
probability of living in a bubble universe where $k$ is such that $\Lambda
_{0}(k)\sim O(\Lambda _{\mathrm{obs}})$. Larger values of $\Lambda $ require
smaller $k,$ and hence larger $N$. We define $f(\Lambda )\,\mathrm{d}\Lambda 
$ to be the probability that $\Lambda \in \lbrack \Lambda ,\Lambda +\,%
\mathrm{d}\Lambda ]$ and $f_{N}(N)\,\mathrm{d}N$ is the probability that $%
N\in \lbrack N,N+\,\mathrm{d}\Lambda ]$. Gibbons and Turok (GT) calculated $%
f_{N}(N)=c(N)e^{-3N}$ for single field, slow roll inflation using the
natural measure on classical solutions in GR \cite{Gibbons}; $c(N)$ has a
relatively weak $N$-dependence. Arguably, this should be multiplied by a
volume weighting factor $e^{3N}$ giving $f_{\mathrm{N}}\approx c(N)$. With $%
N(k)=\bar{N}-\ln (k/\bar{k})/2$ (and $\bar{N}>50-62$ for $\bar{k}%
/a_{0}^{2}H_{0}^{2}<0.02$ in realistic models), we find (up to a
normalization factor): 
\begin{equation*}
f(\Lambda )=f_{N}\left( N(K_{0}(\Lambda )\right) \left\vert \,\mathrm{d}\ln
K_{0}(\Lambda )/\,\mathrm{d}\Lambda \right\vert ,
\end{equation*}%
If $f_{\mathrm{N}}(N)\propto e^{-3N}$ then $\Lambda _{\mathrm{obs}}$ is just
inside the 80\% CI on $\Lambda $ from $f(\Lambda )$.

Including Bayesian selection makes the observed $\Lambda $ appear even more
typical and reduces the dependence on $f_{N}(N)$. If $\Lambda $ is too large
the formation of galaxies is greatly suppressed \cite{btip}. This limits
observable values by $\Lambda \lesssim 10^{3}\Lambda _{\mathrm{obs}}$.
Bayesian selection (in the context of a multiverse) is sufficient to explain
why $\Lambda $ is not too large, but whether or not the $\Lambda _{\mathrm{%
obs}}$ is typical is heavily dependent on the unknown relative weighting of
different values of the CC in the multiverse (i.e. the prior distribution,
here represented by $\mu \lbrack \lambda ]$). In our theory, the unknown
weighting $\mu $ is effectively replaced by the calculable prior $f(\Lambda )
$. In the allowed $\Lambda $-range the $N$ changes by $<2.5\%$ and so $%
f(\Lambda )$ depends only weakly on $f_{\mathrm{N}}(N)$. We follow Ref. \cite%
{tegmark} and use the number of galaxies as a proxy for the number of
observers. If $f_{\mathrm{N}}(N)\approx \mathrm{const}$ in the allowed
range, we find that $\Lambda _{\mathrm{obs}}$ lies just out the 68\% CI,
whereas with $f_{N}\propto \exp (-3N)$ it lies just inside it. In either
case, $\Lambda _{\mathrm{obs}}$ is entirely typical in our model.

The `coincidence' of $t_{U}/t_{\Lambda }\sim O(1)$ or $\Omega _{\mathrm{%
\Lambda }}/\Omega _{\mathrm{m}}\sim O(1)$ is also a typical occurrence in
our model. Observations give $R\equiv \ln (t_{\Lambda }/t_{\mathrm{U}%
})\approx 0.35$. We calculate $|R|<0.35$ has a probability of 9-15\%,
depending on $f_{\mathrm{N}}$. For $|R|<\ln 2$ it is 16\textrm{-}25\%.
Bayesian selection with an assumed uniform prior gives $\approx 4\%$ and $%
8.5\%$ respectively. Similarly seeing $\Omega _{\Lambda 0}\in \lbrack
0.2,0.8]$ has a 14-22\% chance in our model, and $6.8\%$ with just Bayesian
selection.  

At any given location and time, the wave function is dominated by a
classical history in which $\Lambda $ takes a single constant value. This
means that, classically, no evolution of $\Lambda $ can be observed. Yet the
history that dominates, and its associated $\Lambda $ value, is different at
different observation times \cite{footnoteTIME}. We see a history with CC, $\Lambda _{1}$. A observer in our past would see a different history with CC $\Lambda _{2}>\Lambda _{1}$. Yet, for
measurements of $\Lambda _{1}$ and $\Lambda _{2}$ to be compared,
information would have to be sent from one history to another. At the level
of classical physics there is no mechanism for this. Observers will only see
a history consistent with the constant $\Lambda $ given by Eq.(\ref{eq:phi}%
). Crucially, this includes registering all previous measurements of $%
\Lambda $ as being consistent with $\Lambda =\Lambda _{1}$. Put simply, we
do not see the past as an observer in the past would have seen it. This behaviour
implies a new view of time in which the whole history changes slowly. It arises as a direct consequence of having taken $\mathcal{M}$ to be the observer's causal past which in turn was necessary to preserve causality when $\lambda $ was promoted from an external parameter to a field.

As this behaviour  is an integral part of our model, it is tested indirectly
through the $\Omega _{\mathrm{k0}} =-0.0055(2\zeta_{\rm b})$ prediction. Classically, this
movement from one history to another has no directly detectable
consequences. From a quantum perspective, the wave function is dominated by
a superposition of histories with a small spread in $\Lambda $ of $\Delta
\Lambda \sim (\delta ^{2}I_{\mathrm{tot}}/\delta \Lambda ^{2})^{-1/2}$, .
This superposition could give rise to new effects if a system were sensitive
to shifts of $O(\Delta \Lambda )$. However, with $\Omega _{\Lambda 0}\sim
O(1)$, $\Delta \Lambda /\Lambda \sim \Lambda ^{1/2}/M_{\mathrm{pl}}\sim
10^{-60}\ll 1$ today this effect looks undetectably small.

In summary: we have introduced a new approach to solving the cosmological
constant and coincidence problems. The bare CC, $\lambda $, or equivalently
the minimum of the vacuum energy, is allowed to take many possible values in
the wave function, $Z$, of the universe. The value of the effective CC in
the classical history that dominates $Z$ is given by a new equation, Eq.(\ref%
{eq:phi:simp}). This proposal is agnostic about the theory of gravity and
the number of spacetime dimensions. We have applied it in its simplest and
most natural form to a universe in which gravity is described by GR. The
observed classical history will be completely consistent with a non-evolving
cosmological constant. In an homogeneous and isotropic universe with
realistic matter content we find that the observed value of the effective CC
is typical, as is a coincidence between $1/\sqrt{\Lambda }$ and the present
age of the universe, $t_{U}$. Unlike explanations of the CC problem that
rely only on Bayesian selection in a multiverse, our model in independent of
the unknown prior weighting of different $\Lambda $ values, and makes a
specific numerical prediction for the observed spatial curvature parameter, $%
\Omega _{k0}=-0.0055\times (\zeta _{\mathrm{b}}/0.5),$ that is consistent
with current observations but can be tested in the near future. In
conclusion, we have described a new solution of the cosmological constant
problems that is consistent with observations and free of fine-tunings, new
forms of dark energy, or modifications to GR, implies a new view of time and is subject to
high-precision test. 

\acknowledgments DJS acknowledges STFC.


\begin{thebibliography}{99}
\bibitem{footnote1} The sum over matter and metric configurations in $Z[%
\mathcal{M}]$ is performed with the same set of free boundary data, $\left\{
Q^{A}\right\} $, kept fixed as in the definition of $Z_{\Lambda }[M]$.
\bibitem{blackholenote} We neglect black hole boundaries. They make a
negligible contribution to the surface term in Eq.(2).
\bibitem{footnoteINITIAL} The singular nature of $\partial \mathcal{M}_{%
\mathrm{I}}$ mandates an extra surface term on $\partial \mathcal{M}_{I}$%
.See \cite{longpaper} for further details. The initial state and hence all
terms on $\partial \mathcal{M}_{\mathrm{I}}$ are $\lambda $-independent and
so these terms make no contribution to Eq.(\ref{eq:phi:simp}) for $\lambda $.
\bibitem{footnoteTIME} Differently placed observers are most to observe when $t\sim t_{\rm ms}$, where $t_{\rm ms} \sim O(10^{10}\,{\rm yrs})$ is the main-sequence lifetime; they will all most likely observe $\Lambda\sim O(t_{\rm ms}^{-2})$ in our model.
\end{thebibliography}
\end{document}